\documentclass[
reprint,           
superscriptaddress,
amsmath,           
amssymb,           
aps,               
prl,               
notitlepage,       
longbibliography,  
floatfix,          
nofootinbib,
]{revtex4-1}

\usepackage{tensor}     
\usepackage{graphicx}   
\usepackage[
colorlinks=true,        
citecolor=blue,         
linkcolor=blue,         
urlcolor=blue           
]{hyperref}             
\usepackage{bm}         
\usepackage{xcolor}     
\usepackage{array}      
\usepackage{lipsum}
\usepackage{color}      
\usepackage[utf8]{inputenc} 
\usepackage[section]{placeins} 

\newcommand{\Eq}[1]{Eq.~\eqref{#1}}     
\newcommand{\Fig}[1]{Fig.~\ref{#1}}     
\newcommand{\Table}[1]{Table~\ref{#1}}  
\newcommand{\dif}{\mathrm{d}}

\begin{document}
	
\title{Implications for Cosmic Domain Walls from LIGO-Virgo First Three Observing Runs}
	
\author{Yang Jiang}
\email{jiangyang@itp.ac.cn} 
\affiliation{CAS Key Laboratory of Theoretical Physics, 
		Institute of Theoretical Physics, Chinese Academy of Sciences,
		Beijing 100190, China}
\affiliation{School of Physical Sciences, 
		University of Chinese Academy of Sciences, 
		No. 19A Yuquan Road, Beijing 100049, China}
	
\author{Qing-Guo Huang}
\email{Corresponding author: huangqg@itp.ac.cn}
\affiliation{CAS Key Laboratory of Theoretical Physics, 
		Institute of Theoretical Physics, Chinese Academy of Sciences,
		Beijing 100190, China}
\affiliation{School of Physical Sciences, 
		University of Chinese Academy of Sciences, 
		No. 19A Yuquan Road, Beijing 100049, China}
\affiliation{School of Fundamental Physics and Mathematical Sciences
		Hangzhou Institute for Advanced Study, UCAS, Hangzhou 310024, China}

\date{\today}
	
\begin{abstract}
We put constraints on the normalized energy density in gravitaional waves from  cosmic domain walls (DWs) by searching for the stochastic gravitational-wave background (SGWB) in the data of Advanced LIGO and Virgo's first three observing runs. By adopting a phenomenological broken power-law model, we obtain the upper limit of normalized energy density of SGWB generated by DWs in the peak frequency band $10\sim200$ Hz, and get the most stringent limitation at the peak frequency $f_*=35$ Hz, namely  $\Omega_\text{DW}(f_*=35\,\text{Hz})<1.4\times10^{-8}$ at $95\%$ confidence level (CL). Subsequently, we work out the constraints on the parameter space in the appealing realization of DW structure -- the heavy axion model which can avoid the so-called quality problem. 
\end{abstract}
	
\maketitle
{\it Introduction. }
The successful observation of gravitational-wave (GW) signal from compact binary coalescence \cite{LIGOScientific:2021djp} by Advanced LIGO \cite{LIGOScientific:2014pky} and Advanced Virgo \cite{VIRGO:2014yos} has open up a new era of GW astronomy. The superposition of a large number of not loud enough GW incidences can produce a stochastic gravitational-wave background (SGWB) which encodes information about the physics of astrophysical sources. Until the O3 observing period, the LIGO and Virgo Scientific Collaboration has not detected such a SGWB, and therefore they set the upper limits of SGWB \cite{KAGRA:2021kbb}.

Besides the origin of astronomy like compact binary coalescence, SGWB can also have the cosmological origins, including phase transitions \cite{KIBBLE1980183,PhysRevD.30.272,10.1093/mnras/218.4.629,Mazumdar:2018dfl}, primordial perturbations during inflation \cite{1979JETPL30682S,PhysRevD.55.R435,PhysRevD.50.1157}, cosmic string \cite{Kibble_1976,SARANGI2002185,PhysRevD.71.063510,PhysRevLett.98.111101,LIGOScientific:2021nrg} etc. Detecting SGWB signals from these cosmological origins is of great significance for understanding the history of the early Universe. In this letter, we focus on SGWB signal from  cosmic domain wall (DW) \cite{1982PhRvL..48.1867V,PhysRevLett.48.1156} network -- laminar topological defects that form when a discrete symmetry is spontaneously broken. 
Such discrete symmetries arise in various particle physics frameworks beyond the Standard Model (SM), such as Higgs models \cite{Battye:2020jeu,Chacko:2005pe,DiLuzio:2019wsw}, grand unification \cite{LAZARIDES1982157,EVERETT198243}, supersymmetry \cite{Abel:1995wk,Dvali:1996xe,Kovner:1997ca}, non-Abelian discrete symmetries \cite{Riva:2010jm}, axion models \cite{PhysRevLett.48.1156,GrillidiCortona:2015jxo} and so on. Once the breaking of symmetry occurs after inflation, DWs begin to form  and leave different
Hubble patches in different degenerate vacua. Although the formation of DW networks cause potential conflict with cosmological observations if they dominate the energy and then overclose the universe \cite{Zeldovich1974CosmologicalCO}, one can expect that the discrete symmetry is only approximate \cite{Larsson:1996sp} and the energy bias of the potential can induce the annihilation of the DWs. Along with the motion and annihilation of DWs, GWs are radiated and form the SGWB for present observers. Through the observation and investigation of these signals, it is possible to deduce constraints on related theoretical models and explore the physics at the energy scales inaccessible by accelerator laboratories.

An appealing mechanism forming such a DW network is the spontaneous breaking of $U(1)$ Peccei-Quinn (PQ) symmetry  \cite{Peccei1977,Peccei1977ConstraintsIB}. The resulting QCD axion \cite{PhysRevLett.40.223} explains the undetectable CP violation in strong interaction. However, PQ mechanism depends particularly on the axion potential from QCD instantons and it is susceptible to additional global symmetry breaking. It is widely accepted that there is no continuous global symmetry in quantum gravity \cite{Banks:2010zn} and thus PQ solution is spoiled. This so-called quality problem has always been perplexing. To avoid the problem, heavy axion \cite{HOLDOM1982397,HOLDOM1985316,Rubakov:1997vp,Dimopoulos:2016lvn,Gherghetta:2020ofz,ZambujalFerreira:2021cte} is taken to be consideration. Its heavy sector provides a larger contribution to the axion potential and lift the quality of PQ symmetry. In general, the heavy axion decays more easily and leave few detectable objects. Therefore, detecting GW radiation becomes an important way to understand this model.

In this letter, we provide a first search for the SGWB from DWs using the data of Advanced LIGO and Advanced Virgo's first three observing runs. First of all, we take a model-independent analysis and figure out the constraints on the  parameters characterizing the SGWB from DWs. Then, like \cite{ZambujalFerreira:2021cte}, we focus on the heavy axion \cite{HOLDOM1982397,HOLDOM1985316} for explaining strong CP problems and avoiding the vulnerable PQ mechanism.  
In the standard QCD axion model, the DW network makes up a very tiny fraction of energy density of the Universe and leave SGWB undetectable \cite{Hiramatsu:2012sc}. But GWs from the DW network in the heavy axion model can be much stronger because of the large tension carried by heavy axion DWs. 

Recently, based on the evidence for a stochastic common-spectrum process in Pulsar Timing Arrays (PTAs) dataset \cite{NANOGrav:2020bcs,Chen:2021ncc,Chen:2021wdo,Goncharov:2021oub,Chen:2021rqp,Antoniadis:2022pcn}, the energy spectrum of SGWB from DWs has been searched and used to put constraint on axionlike particles in \cite{Ferreira:2022zzo}. In our analysis, instead, the data from  Advanced LIGO and Advanced Virgo's first three observing runs is adopted to probe the mechanics of DWs at much higher energy scale because the frequency band of LIGO and Virgo is much higher than that of PTA.


\bigskip
{\it SGWB from DWs. }
The energy spectrum of SGWB normalized by the critical density of Universe is defined as 
\begin{equation}
    \Omega_\text{gw}(f)=\frac{1}{\rho_\text{c}}\frac{\dif\rho_\text{gw}}{\dif \ln f}, 
\end{equation}
where $\rho_\text{c}$ is present critical density, $\rho_\text{gw}$ is the energy density of GWs, and $f$ is the frequency of GWs. 
In the absence of significant friction from surrounding plasma, the DW network quickly achieves a scaling regime \cite{Kibble_1976,KIBBLE1980183} with the energy density $\rho_\text{DW}=c\sigma H$, where $c\sim\mathcal{O}(1)$ is a model-dependent prefactor, $\sigma$ is the tension or surface energy density of DW, and $H$ is the Hubble rate. This corresponds to the fractional energy density:
\begin{equation}
\alpha =\frac{\rho_\text{DW}}{3H^2M_p^2}, \label{eq:alpha}
\end{equation}
where $M_p=(8\pi G)^{-1/2}$ is the reduced Planck energy scale. Notice that DWs can possess large energy and become dangerous to cosmology \cite{Zeldovich1974CosmologicalCO,1975JETP401Z}. Therefore, a mechanism inducing the annihilation of DW network at certain time is usually needed in practice. 

 Due to the time-varying quadrupole, DW network emits GWs and the simple estimations show that $\rho_\text{gw}\sim\rho_\text{DW}^2/(32\pi H^2M_p^2)$ \cite{PhysRevD.23.852,Kawasaki:2011vv,Hiramatsu:2013qaa,PRESKILL1991207,Chang:1998tb,Gleiser:1998na}. Most GWs are radiated at the frequency $\tilde{f}_*\simeq H$ corresponding to the time of annihilation and then redshifted to today with the peak frequency:
 \begin{equation}
     f_* \simeq 1.5\left(\frac{g_*}{106.75}\right)^{1/6}\left(\frac{T_*}{10^7\,\text{GeV}}\right)\,\text{Hz}, \label{eq:f_star}
 \end{equation}
where $g_*$ is the number of relativistic degrees of freedom at annihilation. We take the number of (entropy) relativistic degrees of freedom $g_{*,s}=g_*=106.75$ in the whole letter. By assuming annihilation happens at radiation domination period, numerical simulations \cite{Hiramatsu:2012sc,Hiramatsu:2013qaa} show that the normalized energy density of SGWB takes the form 
\begin{equation}
    \Omega_\text{DW}(f) h^2\simeq 4.7\times10^{-7}\tilde{\epsilon}\alpha_*^2\left(\frac{106.75}{g_*}\right)^{1/3}\mathcal{S}\left(\frac{f}{f_*}\right). \label{eq:gwspectrum}
\end{equation}
Here $\alpha_*$ to represent the fractional energy density at annihilation. $\tilde{\epsilon}\simeq 0.1-1$ is a numerical efficiency factor which is suggested in \cite{Hiramatsu:2013qaa}, and its value is fixed to be $0.7$ in our analysis. The causality ensures $\Omega_\text{DW}\sim f^3$ when $f<f_*$. Besides, $\Omega_\text{DW}\sim f^{-1}$ when $f\to\infty$ until the cutoff depicted by the inverse of wall width. The order of $\mathcal{O}(1)$ width around the maximum
is suggested by numerical studies \cite{Hiramatsu:2013qaa}. Therefore, the shape of SGWB spectrum can be parameterized as a broken power law: 
\begin{equation}
    \mathcal{S}(x)=\frac{4}{x^{-3}+3x}.
\end{equation}

\begin{figure}[ht]
    \centering
    \includegraphics[width=0.9\columnwidth]{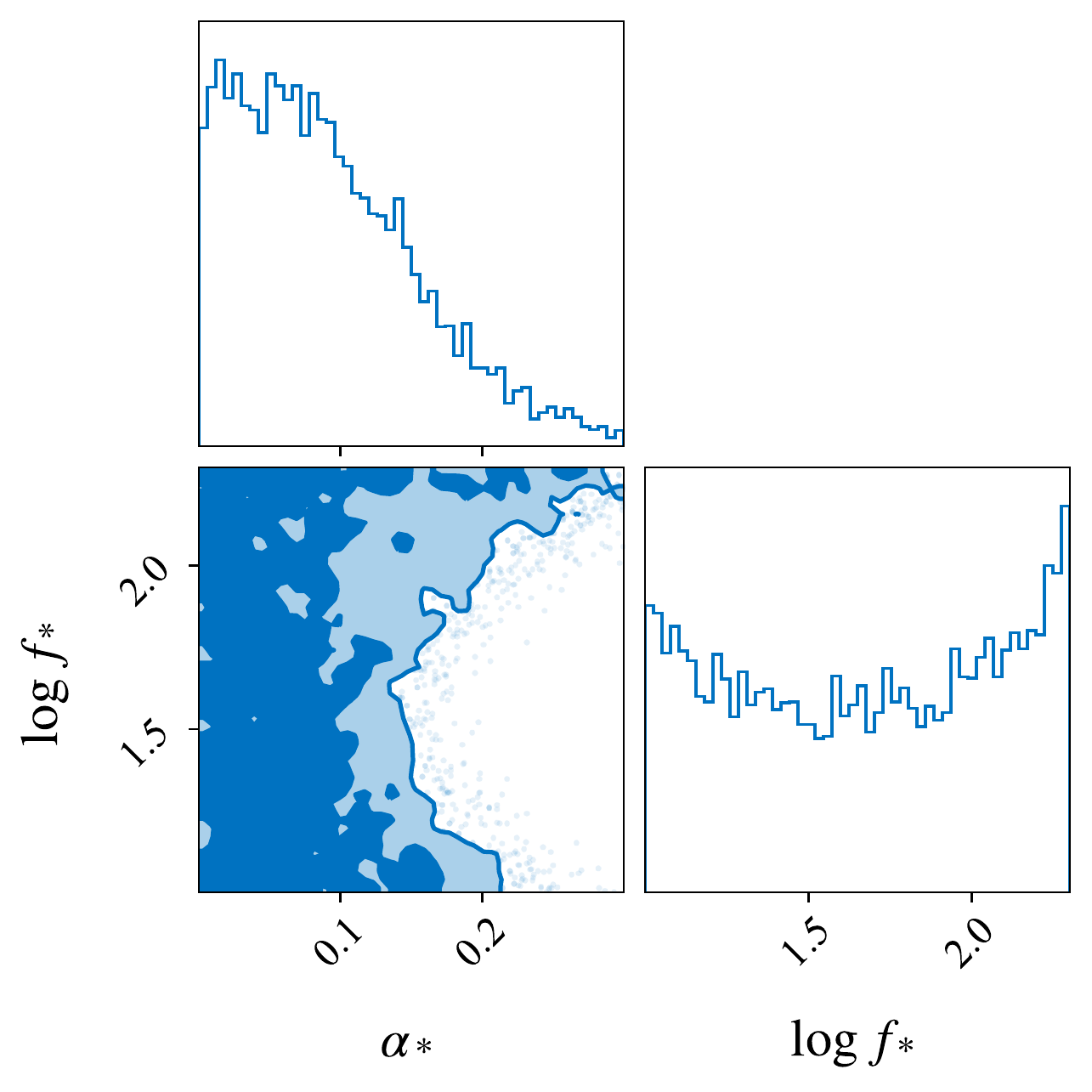}
    \caption{Posterior distribution of $\alpha_*$ and $f_*$ for the broken power law model. The $68\%$ and $95\%$ CL. exclusion contours are shown with blue shaded region.}
    \label{fig:alphafBPL}
\end{figure}

\begin{figure}[ht]
    \centering
    \includegraphics[width=\columnwidth]{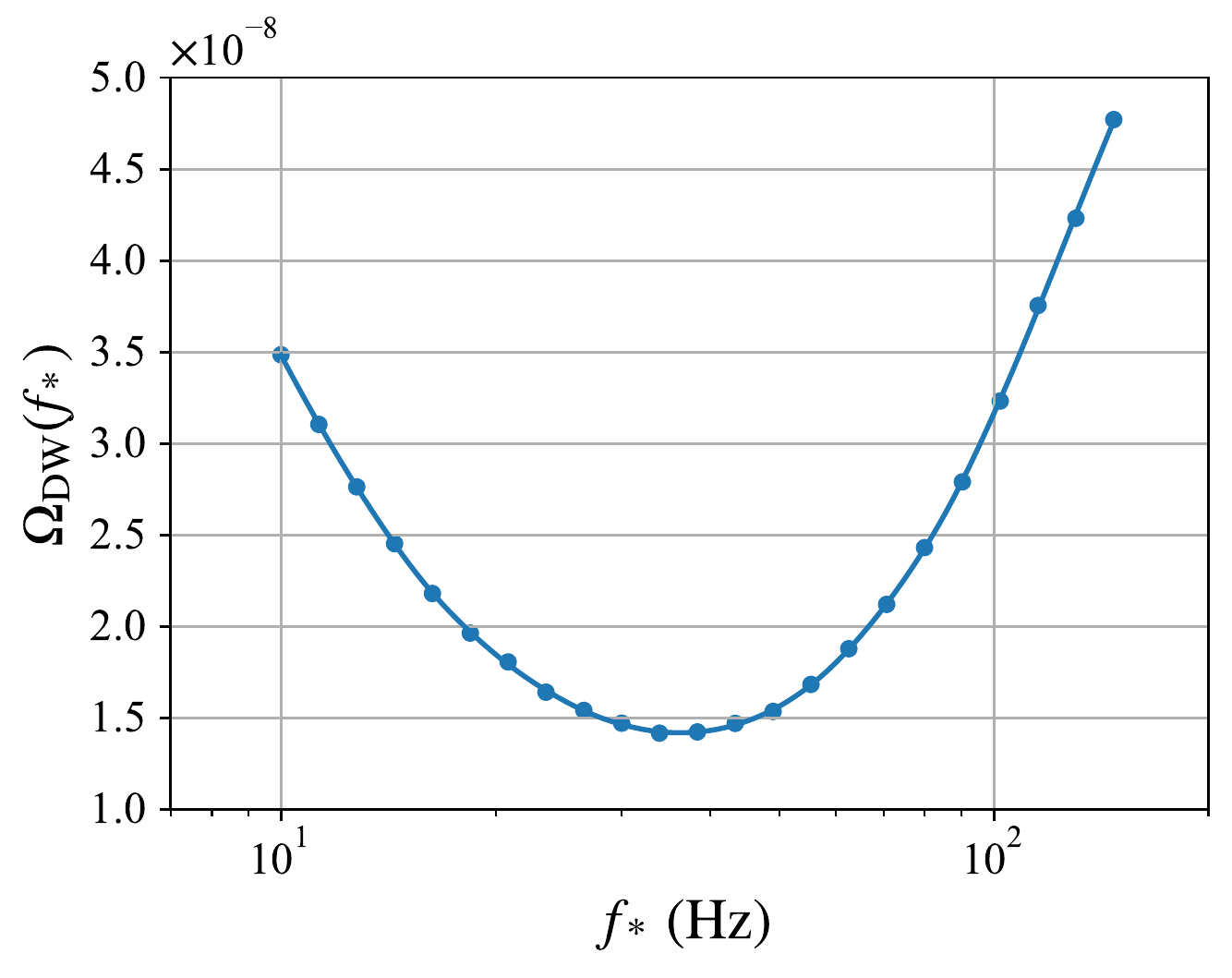}
    \caption{Upper limit of $\Omega_\text{DW}(f_*)$ at 95\% CL for different peak frequency $f_*$.}
    \label{fig:omega}
\end{figure}

We adopt cross-correlation spectrum of isotropic SGWB during advanced LIGO's and advanced Virgo's O1$\sim$O3 observing runs \cite{KAGRA:2021kbb} and follow the method described in \cite{PhysRevLett.109.171102,Romano:2016dpx} to do Bayesian analysis and estimate the probability of models. The likelihood is given by 
\begin{equation}
    p(\hat{C}|\bm{\theta};\lambda)\propto \exp\left[-\frac12\sum_{IJ}\sum_{f}\frac{\left(\hat{C}_{IJ}(f)-\Omega_\text{gw}(f;\bm{\theta})\right)^2}{\sigma^2_{IJ}(f)}\right]. \label{eq:likelihood}
\end{equation}
The model of energy spectrum $\Omega_\text{gw}(f;\bm{\theta})$ is parameterized by a series parameters $\bm{\theta}$. $\hat{C}_{IJ}$ is the cross-correlation spectrum and $\sigma_{IJ}$ denotes its variance. Likelihoods obtained by each detector pair are multiplied together to obtain \Eq{eq:likelihood}. The Bayes factor between SGWB signal and pure noise is used to show the fitness of the model. 
Here $\bm{\theta}=(\alpha_*,f_*)$ are free parameters and their priors are listed in \Table{tab:priorBPL}.
 \begin{table}[ht]
    \centering
    \begin{tabular}{p{0.3\columnwidth}<{\centering} p{0.5\columnwidth}<{\centering}}
        \hline\hline
         Parameter & Prior \\
         \hline
         $\alpha_*$ &  $\text{Uniform}\:[0,\,0.3]$ \\
         $f_*\,\text{(Hz)}$ & $\text{LogUniform}\:[10 ,\,200]$ \\
         \hline\hline
    \end{tabular}
    \caption{Priors of the parameters adopted in broken power law model.}
    \label{tab:priorBPL}
\end{table}
The Bayesian analysis makes use of python {\it bilby} package \cite{Ashton_2019} and dynamic nested sampling package {\it dynesty} \cite{sergey_koposov_2022_6609296}. The posterior distributions of parameters in the broken power law model is presented in \Fig{fig:alphafBPL}. The Bayes factor is $\log\mathcal{B}_\text{noise}^\text{DW}=-0.27$, indicating that no evidence of such signal in the strain data. 
In fact, the posterior of $\alpha_*$ allows us to put constraint on the amplitude of the energy spectrum of GWs from DW network. In \Fig{fig:omega}, the upper limit of $\Omega_\text{DW}(f_*)$ at $95\%$ CL is illustrated for different peak frequency $f_*$. The most stringent limitation is  $\Omega_\text{DW}(f_*=35\,\text{Hz})<1.4\times10^{-8}$ at $95\%$ CL.

\begin{figure}[ht]
    \centering
    \includegraphics[width=0.9\columnwidth]{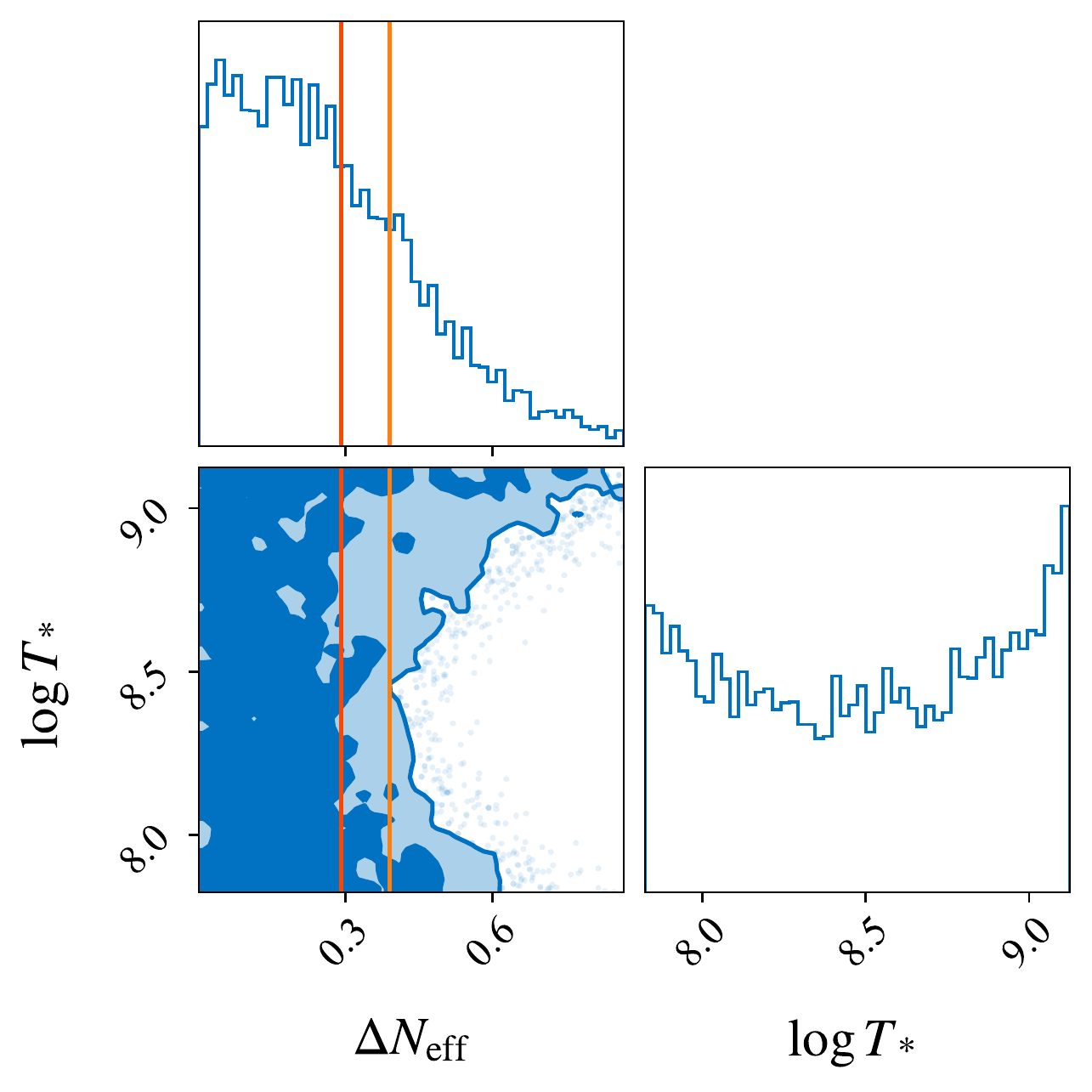}
    \caption{Posterior distribution of $\Delta N_\text{eff}$ and $T_*$ (GeV) for the broken power law model. The $68\%$ and $95\%$ CL. exclusion contours are shown with blue shaded region. The vertical lines denote the constraint from BBN (CMB+BAO).}
    \label{fig:NeffTBPL}
\end{figure}

After the annihilation, DWs may decay into SM particles or dark radiation (DR). Usually the abundance of DR is described by the effective number
of neutrino species $\Delta N_\text{eff}:=\rho_\text{DR}/\rho_\nu$. 
If DWs decay into DR entirely,  $\rho_{DR}\simeq\rho_\text{DW}$ and then 
\begin{equation}
    \Delta N_\text{eff,decay}\simeq 13.6 g_*^{-1/3} \alpha_*,  \label{eq:Neff}
\end{equation}
In addition, GWs emitted by the DW network also make a contribution to DR, namely 
\begin{equation}
    \Delta N_\mathrm{eff,gw}\simeq 0.09\alpha_*^2\left(\frac{g_*}{106.75}\right)^{-1/3}. \label{eq:Neff_gw}
\end{equation}
In this scenario, $\Delta N_\mathrm{eff}=\Delta N_\mathrm{eff,decay}+\Delta N_\mathrm{eff,gw}$. 
It is worth mentioning that BBN (CMB+BAO) also put constraints on the abundance of DR with $\Delta N_\text{eff}\leq0.39$ \cite{Fields:2019pfx} ($0.29$ \cite{Planck:2018vyg}) at $95\%$ confidence level (CL). 
This may also provide a constraint on the energy density of DW network. 
In \Fig{fig:NeffTBPL}, we transform the posterior of $(\alpha_*,f_*)$ to $(\Delta N_\mathrm{eff},T_*)$ using Eqs.~(\ref{eq:f_star}), (\ref{eq:Neff}) and  (\ref{eq:Neff_gw}). It implies that the constraints from LIGO-Virgo is comparable to those from BBN and CMB+BAO in the scenario where DWs are all assumed to decay into DR. 

\bigskip
{\it Heavy QCD Axion Model. } From now on, we will use the former results to constrain the parameter space in the heavy axion model. The composite potential of heavy axion periodic potential and the contribution from QCD is given by 
\begin{equation}
    V_a=\left(\kappa_\text{Q}^2\Lambda_\text{Q}^4+\kappa_\text{H}^2\Lambda_\text{H}^4\right)\left[1-\cos\left(\frac a{F_a}\right)\right], \label{eq:V_a}
\end{equation}
where $F_a$ is axion decay constant, $\Lambda_\text{Q,H}$ are coupling scales of QCD and heavy sector. The factors $\kappa_\text{Q,H}\leq1$ are related to mass of fermion. After the spontaneous breaking of $U(1)$ PQ symmetry, axionic string-DW network forms and then GWs are emitted. In the form of \Eq{eq:V_a}, periodic potential from the heavy sector is aligned with that of QCD. It is typically ensured by the $\mathbb{Z}_2$ symmetry \cite{Hook:2019qoh}. To that end, QCD cannot induce the annihilation of DWs. However, regardless of its specific physical origin (See some models in \cite{Dvali:2005an,DINE1986109,Svrcek:2006yi,Kamionkowski:1992mf}), we could expect a term misaligned with $V_a$ as follows 
\begin{equation}
    V_b\simeq-\mu_b^4\cos\left(\frac{N_b}{N_\text{DW}}\frac{a}{F_a}-\delta\right) \label{eq:V_b}, 
\end{equation}
    where $\mu_b$ is related to the specific model under consideration, $N_b,\ N_\text{DW}$ are integers describing discrete symmetry subgroup. Note that energy bias of potential only appears when $N_b=1$ or co-prime with $N_\text{DW}$. $\delta$ is the generic misaligned phase. CP violation caused by $V_b$ is depicted by
\begin{equation}
    \Delta\theta =\theta-\theta_\text{Q}\simeq r^4\left(\frac{N_b}{N_\text{DW}}\right)\left(\frac{\sin\delta}{\kappa_H^2}\right), \label{eq:CPviolation}
\end{equation}
where $r=\mu_b/\Lambda_\text{H}$. Once $N_\text{DW}>1$, a long-lasting DW network could form, and the fractional energy density is
\begin{equation}
    \alpha_*\simeq0.1\left(\frac{\sqrt{d}c\kappa_\text{H}}{\sin(N_b\pi/N_\text{DW})}\right)^2\left(\frac{F_a}{10^{12}\,\text{GeV}}\right)^2\left(\frac{0.002}{r}\right)^4, \label{eq:alpha_axion}
\end{equation}
and the temperature at annihilation reads 
\begin{multline}
    T_*\simeq 10^8\frac{\sin(N_b\pi/N_\text{DW})}{\sqrt{dc\kappa_\text{H}}}\left(\frac{106.75}{g_*}\right)^{1/4}\left(\frac{r}{0.005}\right)^2 \\ \left(\frac{10^{12}\,\text{GeV}}{F_a}\right)^{1/2}\left(\frac{\Lambda_\text{H}}{10^{10}\,\text{GeV}}\right)\,\text{GeV}.\label{eq:Tstar_sxion}
\end{multline}

Different from the standard QCD axion, the heavy axion tends to decay into SM particles: gluons, photons or fermions, depending on the energy scale, and then the heavy axion avoids the constraints on the abundance of DR from BBN and CMB+BAO. If the heavy axion decays not fast enough, there is a temporary period of matter domination (MD) and hence  \Eq{eq:f_star} and \Eq{eq:gwspectrum} should be modified. To evaluate whether the situation takes place, we compare the characteristic temperature of efficient decay with that of MD, where, according to \cite{ZambujalFerreira:2021cte}, the effective characteristic temperature of decay is
\begin{equation}
    T_\text{d}\simeq 10^6\left(\frac{\sqrt{\kappa_\text{H}}\Lambda_\text{H}}{10^{10}\,\text{GeV}}\right)^3\left(\frac{10^{12}\,\text{GeV}}{F_a}\right)^{5/2}\text{GeV}, \label{eq:T_d}
\end{equation}
and the temperature of MD is 
\begin{multline}
    T_\text{MD}=8\times10^{6}\left(\frac{\sqrt{d}(c\kappa_\text{H})^{3/2}}{\sin(N_b\pi/N_\text{DW})}\right)\left(\frac{F_a}{10^{12}\,\text{GeV}}\right)^{3/2} \\ \left(\frac{\Lambda_\text{H}}{10^{10}\,\text{GeV}}\right)\left(\frac{0.001}{r}\right)^2\left(\frac{106.75}{g_*}\right)^{1/4}\text{GeV}. \label{eq:T_MD}
\end{multline}
If $T_\text{d}<T_\text{MD}$, there is a MD phase, and the peak frequency and the energy spectrum of SGWB from DWs become \cite{ZambujalFerreira:2021cte}, 
\begin{gather}
    f_{*} \to  f_*\left(\frac{T_\text{d}}{T_\text{MD}}\right)^{1/3}, \\
    \Omega_\text{DW} \to \Omega_\text{DW}\left(\frac{g_*(T_\text{d})}{g_*(T_\text{MD})}\right)^{1/3}\left(\frac{T_\text{d}}{T_\text{MD}}\right)^{4/3}.
\end{gather}

In fact, there are also some inherent limitations which should be taken into account in our analysis. Firstly, the characteristic temperature $(T_\text{DW-dom})$ of string-wall dominating the Universe can be estimated by $\alpha_*\simeq1$ and
\begin{multline}
    T_\text{DW-dom}\simeq 5\times10^6\sqrt{\frac{c\kappa_\text{H} F_a}{10^{12}\,\text{GeV}}}\\
    \left(\frac{\Lambda_\text{H}}{10^{10}\,\text{GeV}}\right) \left(\frac{106.75}{g_*}\right)^{1/4}\:\text{GeV}.
\end{multline}
In the case of $T_\text{DW-dom}>T_*$, DWs will dominate our Universe, which may conflict with the standard cosmology. Even if the contradiction might be avoided  \cite{Kawasaki:2004rx}, there is a lack of knowledge about the evolution of DWs in the situation. Secondly, the validity of the axion effective theory requires  $\sqrt{\kappa_\text{H}}\Lambda_\text{H}<F_a$. At last, Neutron Electric Dipole Moment (nEDM) measurements \cite{Abel:2020pzs} require $\Delta\theta\lesssim10^{-10}$.

\begin{figure*}[ht]
    \centering
    \includegraphics[width=0.325\textwidth]{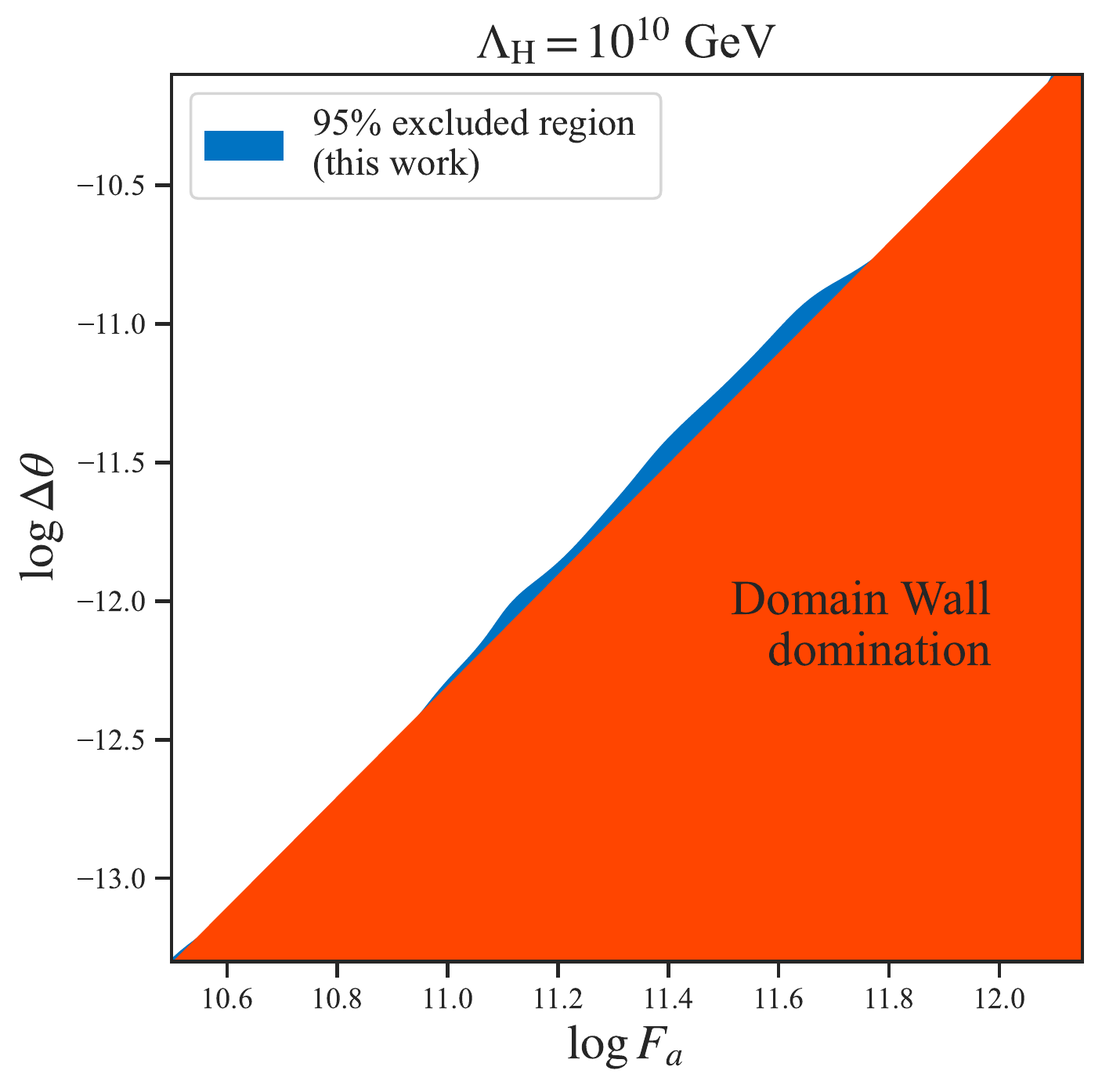}
    \includegraphics[width=0.325\textwidth]{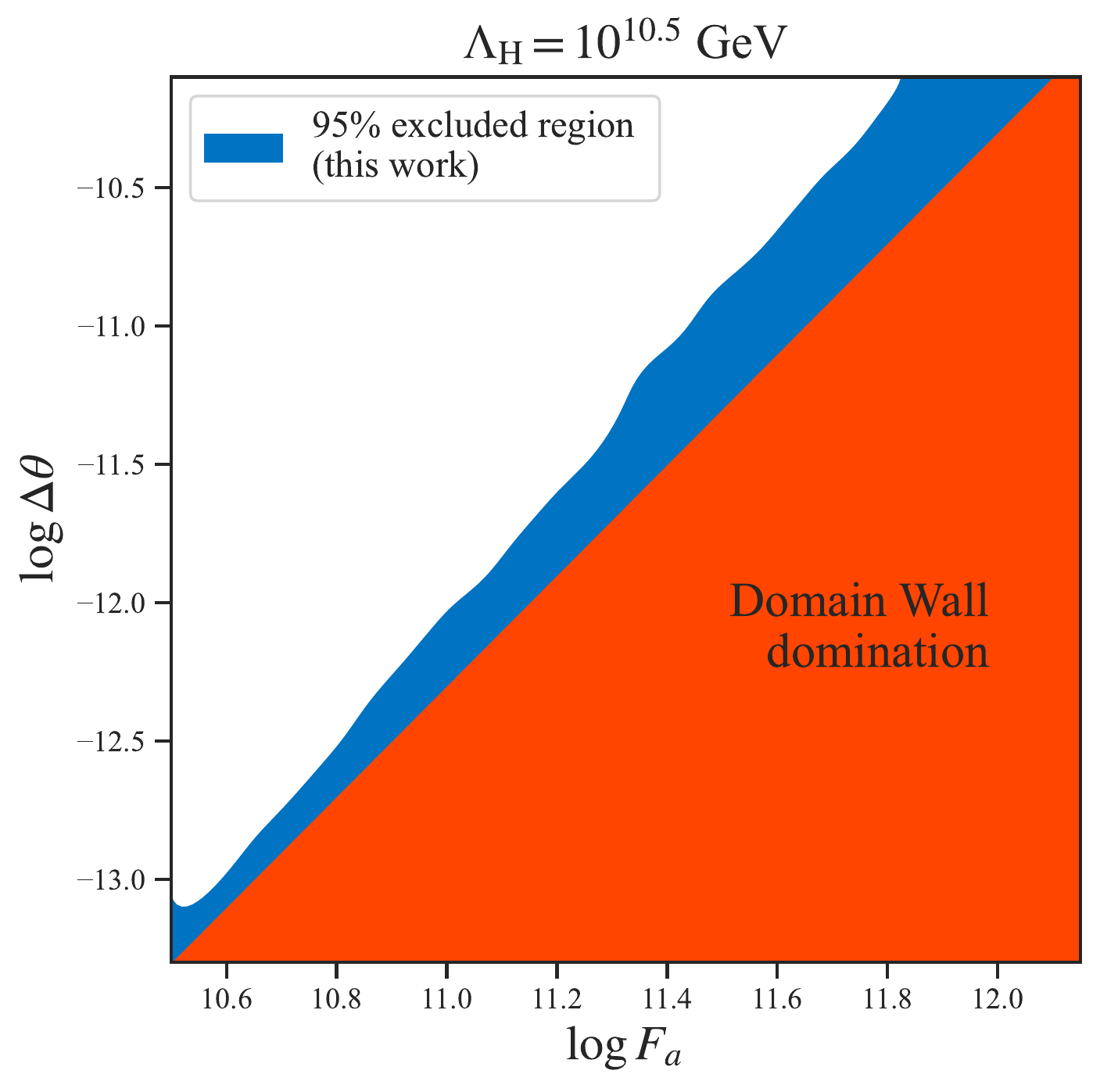}
    \includegraphics[width=0.325\textwidth]{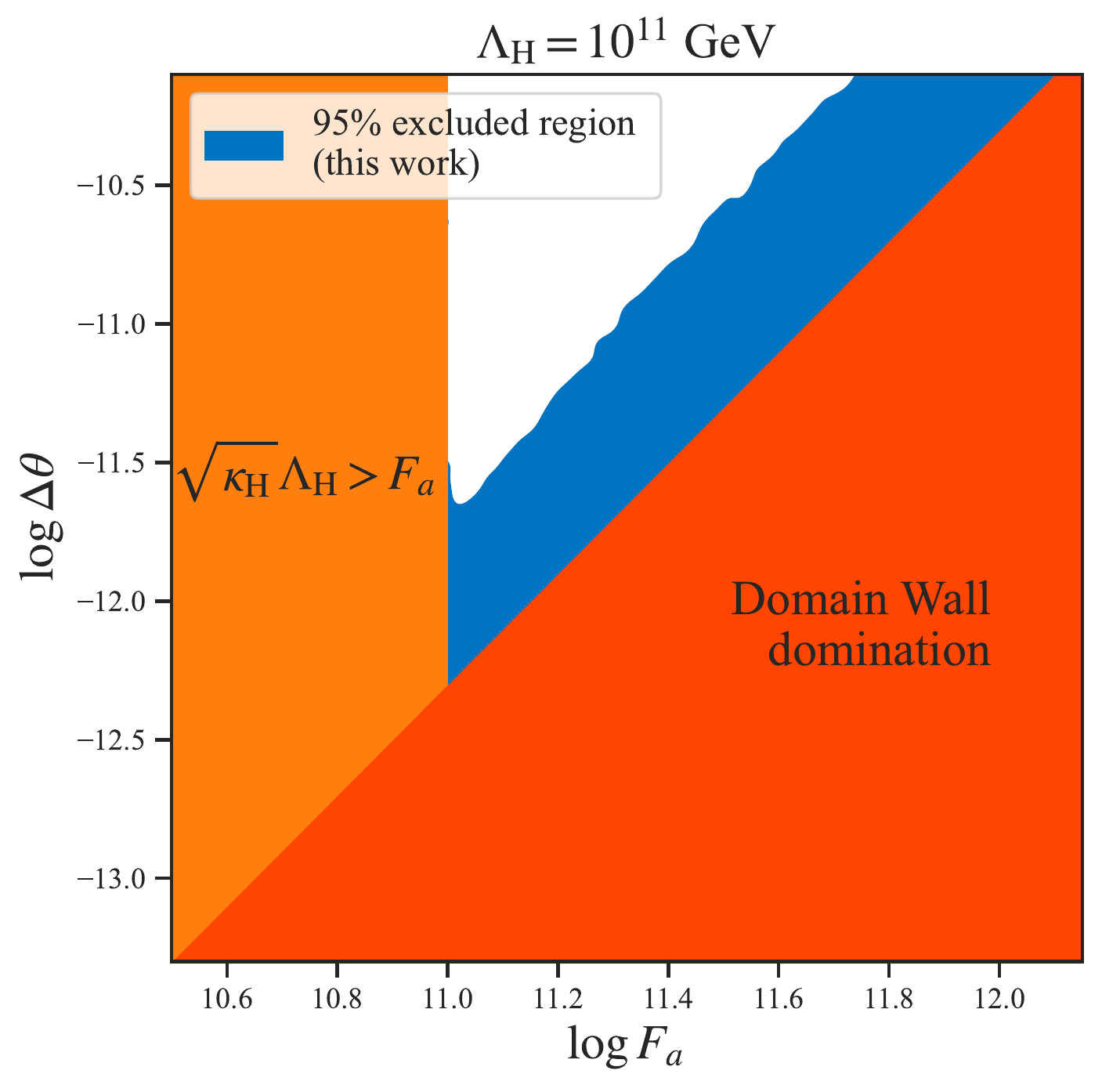} \\ \hspace{0.4em}
    \includegraphics[width=0.325\textwidth]{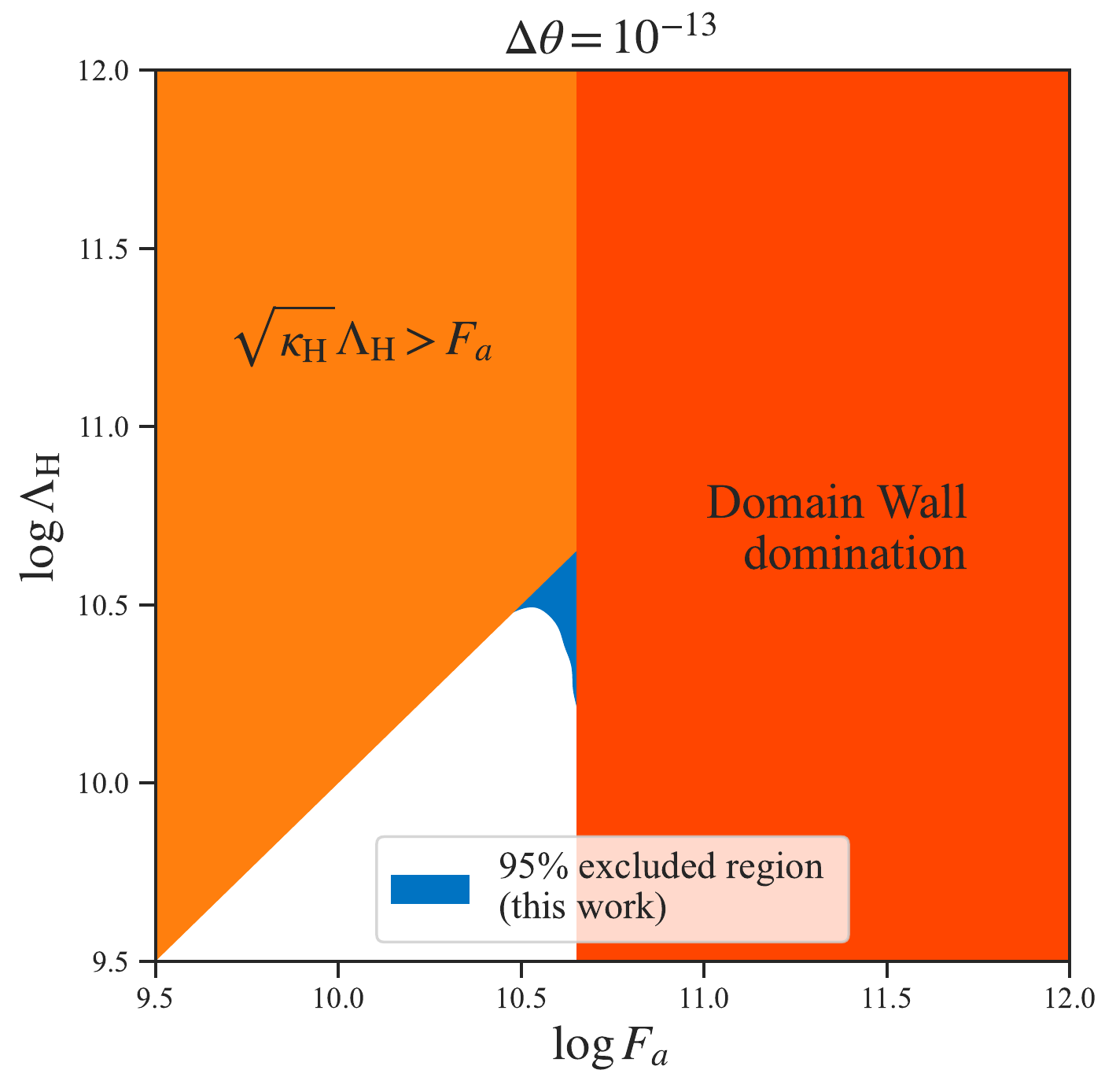}
    \includegraphics[width=0.325\textwidth]{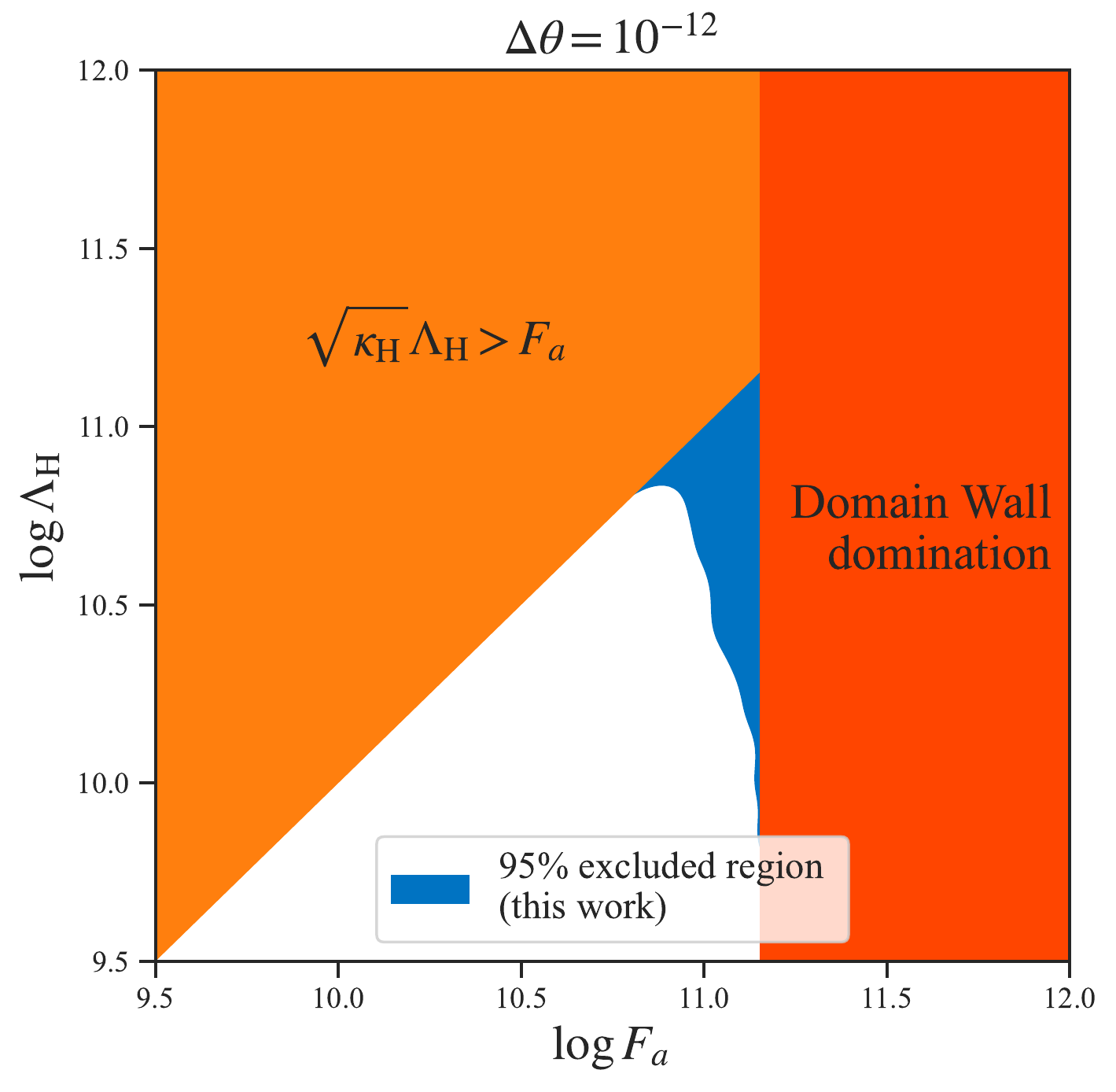}
    \includegraphics[width=0.325\textwidth]{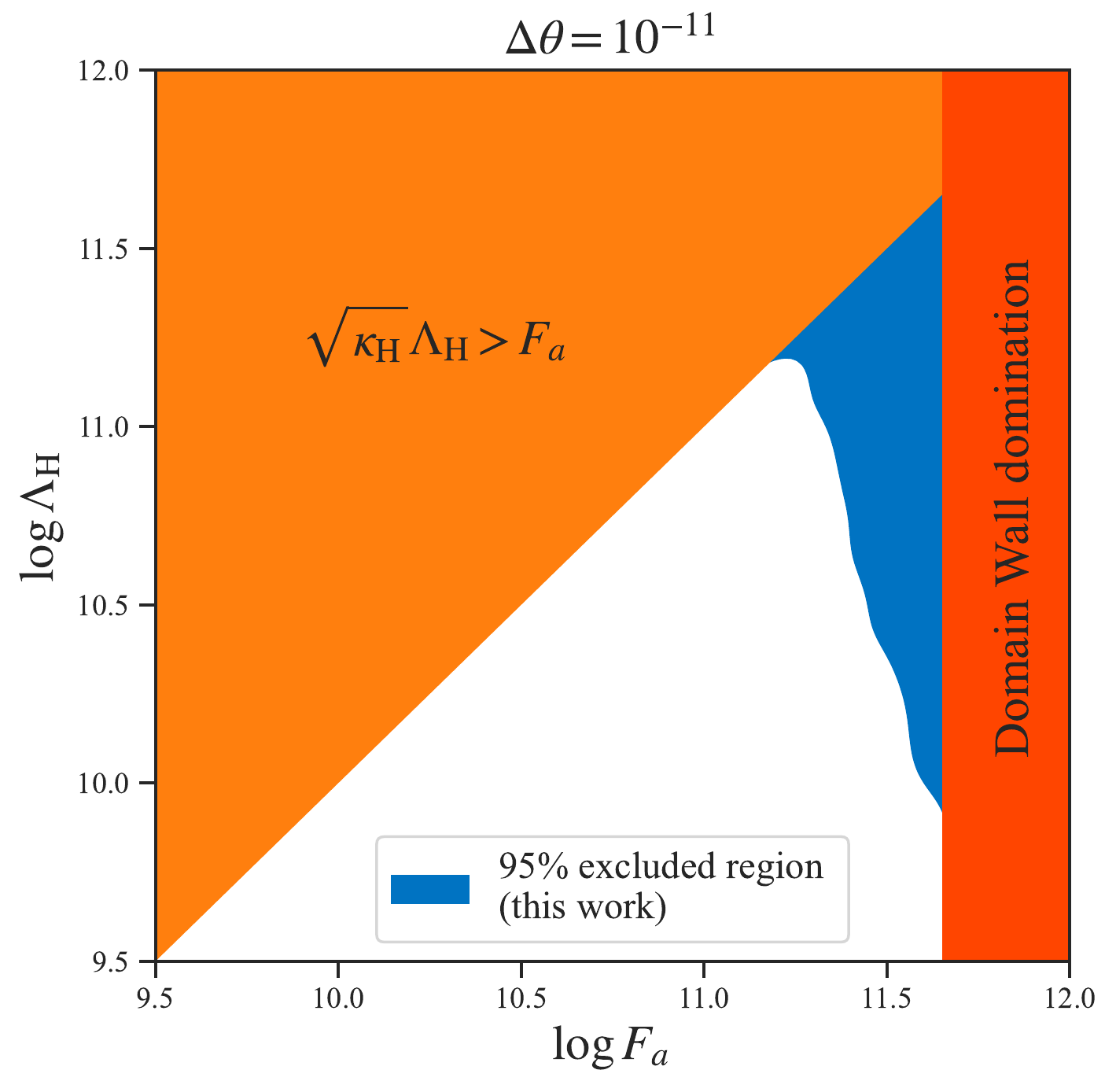}
    \caption{Parameter space  excluded by Advanced LIGO and Advanced Virgo O1$\sim$O3 observing runs. The blue shaded regions denote the 95\% exclusion contours. The orange-red and orange shaded regions correspond to DW domination and violation of axion effective theory, respectively.}
    \label{fig:LambdaThetaFixed}
\end{figure*}

In this letter we mainly focus on constraining the three physical parameters (e.g. the coupling scale $\Lambda_\text{H}$ of heavy sector, axion decay constant $F_a$ and CP violation coefficient $\Delta\theta$) in the heavy axion model. From \Eq{eq:alpha_axion} and \Eq{eq:Tstar_sxion}, the former constraints on $\alpha_*$ and $T_*$ can be used to constrain these three parameters. 
Here we set the factor $\kappa_\text{H}\simeq1$ providing that the heavy sector contains a light fermion. Similar to  \cite{Kawasaki:2014sqa}, we take $N_b=1,N_\text{DW}=6$, and then $c=4.48,\ d=6.28$. The dependence of $\delta$ only appears in \Eq{eq:CPviolation} as a factor between $\Delta\theta$ and $r$. Therefore, without losing generality, for example, we set $\delta=0.3$. Since there are three free parameters, our strategy of analysis is to constrain the other two parameters by keeping $\Lambda_\text{H}$ or  $\Delta\theta$ fixed, respectively. 
The priors adopted in our analysis are listed in \Table{tab:priorAxion}, and the excluded parameter spaces in the heavy axion model are illustrated by the blue shaded regions in  \Fig{fig:LambdaThetaFixed}.  

\begin{table}[ht]
    \centering
    \begin{tabular}{|p{0.3\columnwidth}<{\centering} p{0.5\columnwidth}<{\centering}|}
        \hline
        \multicolumn{2}{|c|}{$\Lambda_\text{H}=10^{10},10^{10.5},10^{11}$ GeV} \\
        \hline\hline
         Parameter & Prior \\
         \hline
         $F_a\,\text{(GeV)}$ &  $\text{LogUniform}\:[10^{10},\,10^{12.2}]$ \\
         $\Delta\theta$ & $\text{LogUniform}\:[10^{-14.5},\,10^{-10}]$ \\
         \hline\hline
        \multicolumn{2}{|c|}{$\Delta\theta=10^{-13},10^{-12},10^{-11}$} \\
        \hline\hline
         Parameter & Prior \\
         \hline
         $F_a\,\text{(GeV)}$ &  $\text{LogUniform}\:[10^{10},\,10^{12.2}]$ \\
         $\Lambda_\text{H}\,\text{(GeV)}$ & $\text{LogUniform}\:[10^{9},\,10^{12.2}]$ \\
         \hline
    \end{tabular}
    \caption{Priors of the parameters adopted in the heavy axion model. 
    }
    \label{tab:priorAxion}
\end{table}

\bigskip
{\it Conclusions.}
By adopting the cross-correlation spectrum of Advanced LIGO and Advanced Virgo's first three observing runs, we search for the GW signal from DW network in the peak frequency range of $10\sim200$ Hz. Since no such signal in the strain data has been observed, we place constraints on the energy spectrum of SGWB from cosmic DW network. The most stringent limitation is $\Omega_\text{DW}(f_*=35\,\text{Hz})<1.4\times10^{-8}$ at $95\%$ CL. The posterior of $f_*$ dose not show obvious preference for any certain peak frequency, and the fractional energy density of DWs with $\alpha_*>0.2$ is excluded at $95\%$ CL for  $f_*=20\sim100$ Hz corresponding to the most sensitive band of LIGO and Virgo.

Furthermore, for fixed values of $\Lambda_\text{H}=10^{10}$,  $10^{10.5}$ or $10^{11}$ GeV, the CP violation coefficient $\Delta \theta$ in the blue shaded regions in the upper panel of \Fig{fig:LambdaThetaFixed} is excluded; for fixed values of $\Delta \theta=10^{-13}$,  $10^{-12}$ or $10^{-11}$, the coupling scale $\Lambda_\text{H}$ of the heavy sector in the blue shaded regions in the bottom panel of \Fig{fig:LambdaThetaFixed} is excluded. In all, even though SGWB has not been detected by Advanced LIGO and Advanced Virgo, the non-detection of the SGWB from cosmic DWs already places  constraints on the heavy axion model.


Our results have shown the ability of LVK interferometers to understand physics related to DWs and we expect more elaborate results with the improvement of the detector sensitivity in the near future.

{\it Acknowledgments. } 
We acknowledge the use of HPC Cluster of ITP-CAS and HPC Cluster of Tianhe II in National Supercomputing Center in Guangzhou. This work is supported by the National Key Research and Development Program of China Grant No.2020YFC2201502, grants from NSFC (grant No. 11975019, 11991052, 12047503), Key Research Program of Frontier Sciences, CAS, Grant NO. ZDBS-LY-7009, CAS Project for Young Scientists in Basic Research YSBR-006, the Key Research Program of the Chinese Academy of Sciences (Grant NO. XDPB15).

\bibliography{ref.bib}
	
\end{document}